# Spin-lattice coupling in the ferrimagnetic semiconductor FeCr$_2$S$_4$ probed by surface acoustic waves

C. Müller[1], V. Zestrea[2], V. Tsurkan[1,2*], S. Horn[1], R. Tidecks[1] and A. Wixforth[1]

[1] *Institut für Physik, Universität Augsburg, D-86159 Augsburg, Germany*
[2] *Institute of Applied Physics, Academy of Sciences of Moldova, MD-2028, Chisinau, R. Moldova*

## ABSTRACT

Using surface acoustic waves, the elastomagnetic coupling could be studied in thin single crystalline plates of the ferrimagnetic semiconductor FeCr$_2$S$_4$ by measuring the attenuation and the frequency tracking in the temperature range 4.2 K to 200 K. The data clearly display the anomalies found in low-field magnetization measurements.

[*] Corresponding author; electronic mail: vladimir.tsurkan@physik.uni-augsburg.de

## I. INTRODUCTION

Spin-lattice coupling in correlated magnetic systems strongly influences the electronic transport and plays an essential role in the formation of the magnetic ground state. For example, electron-phonon interaction and lattice polarons contribute substantially to the colossal magnetoresistance (CMR) effect and the magnetic field induced metal-insulator transition in manganite perovskites.[1-3]

Here, we report on the study of spin-lattice correlations by surface acoustic waves (SAW)[4] in the ternary ferrimagnet $FeCr_2S_4$ with a cubic spinel-type crystal structure at high temperatures, in which the CMR effect was also recently discovered.[5] In this structure the chromium ions occupying the octahedral sites are in a $3d^3$ state with three d-electrons in a lower $t_{2g}$ triplet and an orbital moment quenched by a crystal field. The $Fe^{2+}$ ions occupying the tetrahedral sites are in a $3d^6$ configuration with a hole in a lower e-doublet and thus, are Jahn-Teller (JT) active. Strong on-site interaction of the ferrous ions allows a distortion of the $FeS_4$ tetrahedrons, experimentally detected as local lattice correlation or long-range orbital ordering. The presence of local structural distortions in this compound was originally deduced from the Mössbauer experiments, e.g., the appearance of a quadrupole splitting and anomaly of the electric field gradient induced at the Fe ions sites at a temperature of 10 K.[6,7] These features were explained in the framework of static and dynamic JT effects.[8] An alternative explanation suggested an orbital ordering due to hybridization of Cr and exited Fe states.[9] The interpretation of the Mössbauer data, however, was in conflict with X-ray and neutron scattering diffraction investigations, which state that polycrystalline $FeCr_2S_4$ remains a cubic spinel down to 4.2 K.[10,11] In powdered single crystals, the symmetry was also found to be unchanged, although a broadening of the X-ray diffraction lines was observed. It was attributed to inhomogeneous lattice distortions that develop below the Curie temperature and persist down to ~ 4.2K.[12]

Several recent experimental investigations on $FeCr_2S_4$ single crystals pointed out the importance of a spin-lattice coupling. A cusp-like anomaly in the temperature dependence of the magnetization at $T_m$ ~ 60 K and a splitting of zero-field cooled (ZFC) and field cooled (FC) magnetization below this temperature was observed, which is unexpected for a structurally well ordered ferrimagnet.[13] Hydrostatic pressure investigations[14] show that the magnetic anomaly at $T_m$ in $FeCr_2S_4$ is highly sensitive to lattice contraction. A non-cubic magnetocrystalline anisotropy associated with tetragonal distortions was revealed. AC-susceptibility[15] and magnetoresistance studies [14] attributed the spin-glass-like features to the changes in the magnetic domain structure due to additional pinning centres below $T_m$ as a result of a structural lattice transformation. Later on, ultrasonic measurements of $FeCr_2S_4$ single crystals gave additional evidence for a structural transformation at $T_m$ ~ 60 K. The elastic moduli manifest a step-like feature around this temperature indicating a structural phase transition of first-order type. Below 60 K a pronounced softening of the elastic moduli was detected. The experimental data, however, indicate the appearance of a trigonal distortion, which was explained in terms of an orbital ordering with coupling of the orbitals of Fe ions along the <111> direction.[16] Very recently, high resolution electron-microscopy studies of $FeCr_2S_4$ single crystals[17] have revealed a peculiar structural transformation below 60 K indicating a cubic-to-triclinic symmetry reduction within crystallographic domains. The overall crystal symmetry was found to be reduced from Fd3m to F43m. The triclinic distortions were suggested to result from the combined actions of tetragonal distortions due to the JT active $Fe^{2+}$ ions and trigonal distortions due to a displacement of the $Cr^{3+}$ ions in the <111> direction.

## II. EXPERIMENTAL DETAILS

The $FeCr_2S_4$ single crystals were grown by the chemical transport reaction method,[18] using chlorine as a transport agent. The single phase spinel structure was confirmed by X-ray diffraction analysis of the powdered single crystals. The sample composition was determined by electron-probe microanalysis that found nearly stoichiometric (within 1%) composition and a small amount (1%) of chlorine that substitute the sulphur ions. Samples for the SAW study were cut from the octahedron crystals in form of thin optically polished plates (thickness ~ 20µm, area ~ 1.1 × 1.2 $mm^2$) with different plane orientations, (111) and (100). In a cubic system (like a spinel) the plane normals point into the directions with the same Miller indices.

The samples were coupled to a $LiNbO_3$ delay line.[19] Their plane normals show perpendicular to the surface of the substrate. For an intense mechanical coupling we used diluted GE varnish. Thus, the SAW propagates in the planes of the $FeCr_2S_4$ plates, which extend across the whole width of the sound path. The measurements were done in a cryostat with variable temperature insert (VTI), working in the temperature range from room temperature to 4.2 K.

To generate SAWs, micro-finger structures made of aluminium deposited onto a piezoelectric substrate ($LiNbO_3$ with 128° rotated YX cut) were used,[4,20] as shown in Fig.1. At the contact pads of these so called interdigital transducers (IDTs) a radio frequency voltage with fundamental frequency $f_0$ is applied, which generates by the inverse piezoelectric effect a deformation propagating with sound velocity $v$ over the delay line. We used "split-1-finger electrodes" so that the distance $b$ between two fingers connected to the same pad is equal to $\lambda_{IDT}$, the wavelength of the SAW generated by the IDT. The velocity $v$ of the SAW, its wavelength $\lambda$, and the fundamental frequency $f_0$ are connected via the relation:

$$\lambda \cdot f_0 = v \qquad (1)$$

The IDT emits the SAW in a superimposed way, because after propagating to the next finger pair the deformation is enhanced. The electrical field of the SAW propagating over the delay line (length ~5.8 mm midpoint to midpoint of the IDTs) is detected using a second IDT being identical to the first one. For the radio frequency generation and detection a vector network analyzer (NWA, ZVC Rohde und Schwarz) was used. As the finger distances and the sound veloctiy in LiNbO$_3$ slightly change with temperature, the fundamental frequency which is transmitted with the minimal attenuation, varies and, therefore, has to be tracked. This eliminates the attenuation fraction arising from the filter characteristic of the IDT. The tracking was done by measuring the attenuation of the SAW in a frequency range spanned around the fundamental frequency of the SAW device. Then, the frequency with the lowest attenuation is taken as the fundamental frequency $f_0$ and its attenuation is read out.

The distance $b = \lambda_{IDT}$ of the IDT used was 19.25 µm. The frequency $f_0$ was tracked in the range from ~ 201 MHz (at room temperature) to ~ 204.5 MHz at low temperatures. The sound velocity of 128° rot YX cut (i.e. propagation of the SAW in x-direction) LiNbO$_3$ is $v_0$ = 3978.2m/s at room temperature.[21] The split-1-finger electrodes used, consist of relatively broad metal lines with a small distance between, yielding a metallized-like behaviour of the LiNbO$_3$ in the region of the IDTs. Calculating the velocity from the geometry of the IDT (b= $\lambda_{IDT}$) and the frequency ($f_0$ = 201 MHz) according to eq. (1), therefore, gives $v_{0,IDT}$=3869.3 m/s, which is the value of electrically shorted 128° rot YX cut LiNbO$_3$, obtained by metallizing the substrate by an aluminium film.[21]

Since the frequency $f_0$ of the voltage is impressed to the IDT, the wavelength with which the SAW propagates along the LiNbO$_3$ not covered by the IDT according to eq. (1) is given by $\lambda_0 = v_0/201$ MHz $=19.79$ μm at room temperature.

As the measured attenuation does not only contain the signal of the SAW, but also a contribution from a direct electromagnetic coupling of the two IDTs, a special method had to be applied to distinguish between the direct electrical crosstalk received by the second IDT and the SAW signal. This procedure is described in Ref. 22, which, moreover, contains additional valuable information about the measuring method.

### III. RESULTS AND DISCUSSION

The temperature dependence of the attenuation for the sample with the normal of the (100) plane perpendicular to the substrate and the SAW propagating in the <110> crystallographic direction, is presented in Fig.2. It shows a non-monotonic behaviour with pronounced anomalies at around 170 and 60 K. These features correlate well with the changes of the low-field magnetization presented in the same figure. At the Curie temperature $T_C$, the attenuation manifests a maximum followed by a pronounced decrease down to approximately 40K. At the temperature $T_m$ of the spin-glass like magnetization anomaly, the attenuation shows a sharp peak and irreversible behaviour, resembling that of the hysteretic behaviour of the magnetization. Finally, below 40 K, the attenuation starts to increase again but flattens below ~20K.

Fig. 3 depicts the temperature dependence of the fundamental frequency $f_0$ for the same sample. It exhibits well resolved features at the characteristics temperatures of the system, namely at $T_C$ and $T_m$. Additionally, a less pronounced anomaly in $f_0$ can be also noted at a temperature of 130 K. This temperature corresponds to a minimum of the electrical resistivity of such samples.[14]

Thus, the frequency tracking mirrors the behaviour of the attenuation. To understand this, one has to consider the propagation conditions for the SAW along the sound path from the sending to the receiving IDT. Once emitted by the sending IDT, the SAW propagates according to $\lambda_0 f_0 = v_0$ towards the region of the $FeCr_2S_4$ plate. The plated region represents a "sandwich system" with a sound velocity $v_p$ which is different from the free surface velocity $v_0$, because some other material is put on top of the surface.[4, 21, 23-25] In the region of the $FeCr_2S_4$ plate the wavelength of the SAW adjusts according to $\lambda_p f_0 = v_p$. When propagating into $LiNbO_3$ with a free surface again, the wavelength has again the value $\lambda_0 = v_0/f_0$.

By tracking the frequency to get a best transmitted signal with minimal attenuation, $f_0$ should be adjusted so that the wavelength $\lambda_{IDT}$ fits to the periodicity of the finger electrodes of the IDTs. However, an adjustment of $f_0$ yields also a change of $\lambda_0$ and $\lambda_p$, so that the tracking procedure optimizes at the same time all of the three wavelengths to get a minimal attenuation. Therefore, changes of the physical properties of the $FeCr_2S_4$ plate which influence $v_p$ and, thus, $\lambda_p$ are displayed in the frequency tracking diagram (Fig. 3).

The attenuation in Fig. 2 is in qualitative agreement with the attenuation observed in bulk ultrasound velocity measurements in $FeCr_2S_4$ single crystals, as plotted in Fig. 1b of Ref.16. Different is that the attenuation above the Curie temperature $T_C$ in Fig. 2 does not exhibit the same low values as in the low temperature region. The reason could be, that in the measurements of Ref. 16 only elastic contributions to the damping are detected. Since the SAW in the present work is a wave on a piezoelectric substrate, the conductivity of the plate may be also important for the damping.[22,26,27] Above $T_C$, the conductivity in $FeCr_2S_4$ is higher than at low temperatures,[14] but still only of order 1 $\Omega^{-1}cm^{-1}$, from which one may expect a considerable contribution to the damping.

A coincidence of changes of the sheet conductance of the $FeCr_2S_4$ plate with the shape of the attenuation curve and frequency tracking diagram, respectively, was, however, not found, except for the structure at 130K in Fig. 3. To see this, we used the sheet conductance σ =$d/\rho$ obtained from Fig. 6 of Ref. 14, where the conductivity $\rho$ of a typical $FeCr_2S_4$ single crystalline specimen with thickness $d = 0.02$ cm is shown. In the temperature region in which the anomalies at $T_m$ and $T_C$ are observed, the sheet conductance ranges between 0.005 $\Omega^{-1}$ and 0.045$\Omega^{-1}$.

The frequency tracking curve in Fig. 3 looks very similar to the temperature dependence of the elastic moduli of $FeCr_2S_4$ given in Ref. 16 in the absence of a magnetic field. Since, these moduli are connected to the sound velocity $v_{si}$ via $c_i = \rho v_{si}^2$ with $\rho$= 3.84 g/cm$^3$ the density of $FeCr_2S_4$, Fig. 3 of the present work seems to be a mirror of the sound velocity changes in the $FeCr_2S_4$ plate, and the elastic constant changes, respectively.

## IV. CONCLUSIONS

In the present work, the anomalies observed in the magnetic behaviour of $FeCr_2S_4$ are found in attenuation and frequency tracking curves of SAW measurements. This indicates a strong coupling between spin and lattice degrees of freedom in this compound. Especially, the fact that the magnetic anomaly at $T_m$ =60 K, where $FeCr_2S_4$ shows no remarkable change in the conductivity,[14] is detected by SAW experiments, indicates that it is related to a structural transformation.

The results are in qualitative agreement with our former ultrasonic studies on this compound. Therefore, we demonstrated that surface acoustic waves are a suitable powerful instrument for the investigation of spin-lattice correlation in the magnetic semiconductors.

**ACKNOWLEDGEMENTS**

The authors want to thank J. Ebbecke for supplying the 128° rot YX cut $LiNbO_3$ substrate with split 1 finger electrodes. We acknowledge the support of the SFB 484 by the VTI American Magnetics cryostat and support of US CRDF-MRDA for crystal growth experiments.

**FIGURE CAPTIONS**

FIG. 1. Surface acoustic wave (SAW) delay line with glued $FeCr_2S_4$ plate on the sound path in the middle between the interdigital transducers (IDTs). On the right side one IDT is sketched showing the distance *b* of the two finger electrodes with the same electric potential. NWA: vector network analyzer.

FIG. 2: SAW attenuation for a single crystalline thin plate of $FeCr_2S_4$ on $LiNbO_3$ as a function of the temperature. The full broad line is the average over several measurements for decreasing temperature. Open squares represent an attenuation measurement for increasing temperature.

For comparison the low-field magnetization (measured as described in Ref. 13) of a single crystal of the same butch is shown (FC: field cooled, measured at decreasing temperature; ZFC: zero field cooled, measured at increasing temperature). The thin line is a guide to the eye.

FIG. 3: Frequency $f_0$ of the exciting voltage as a function of temperature for the same $FeCr_2S_4$ single crystalline plate as in Fig. 2, measured at the same run with the attenuation. Symbols represent measured values, the full line is a guide to the eye.

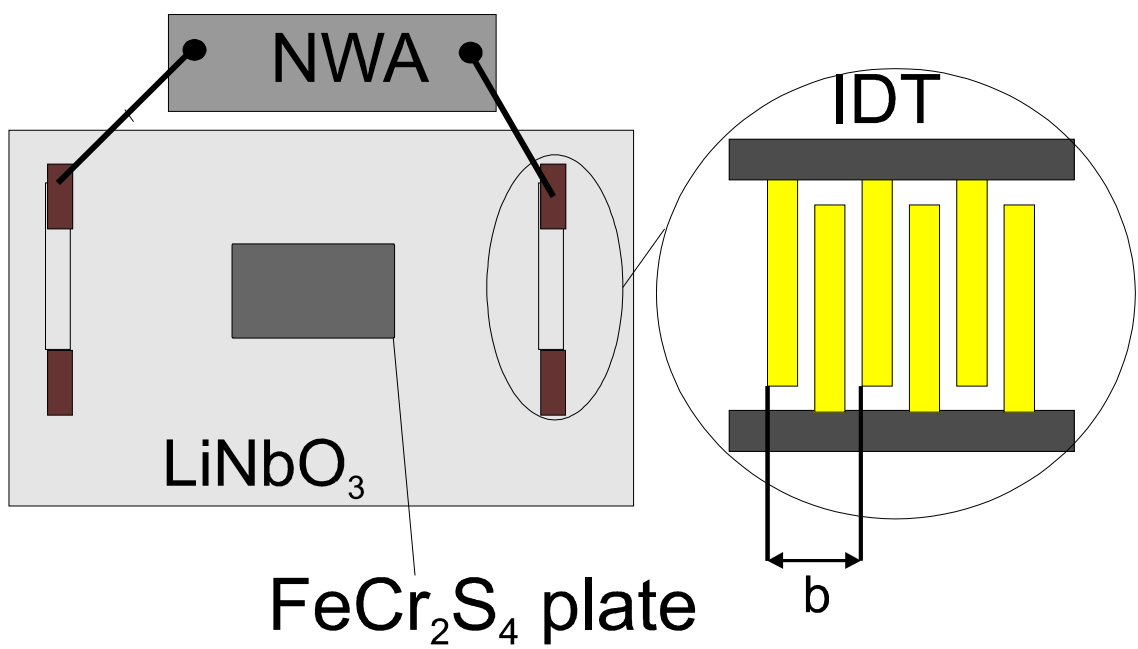

FIG.1  C. Müller , V. Zestrea, V. Tsurkan, et al.

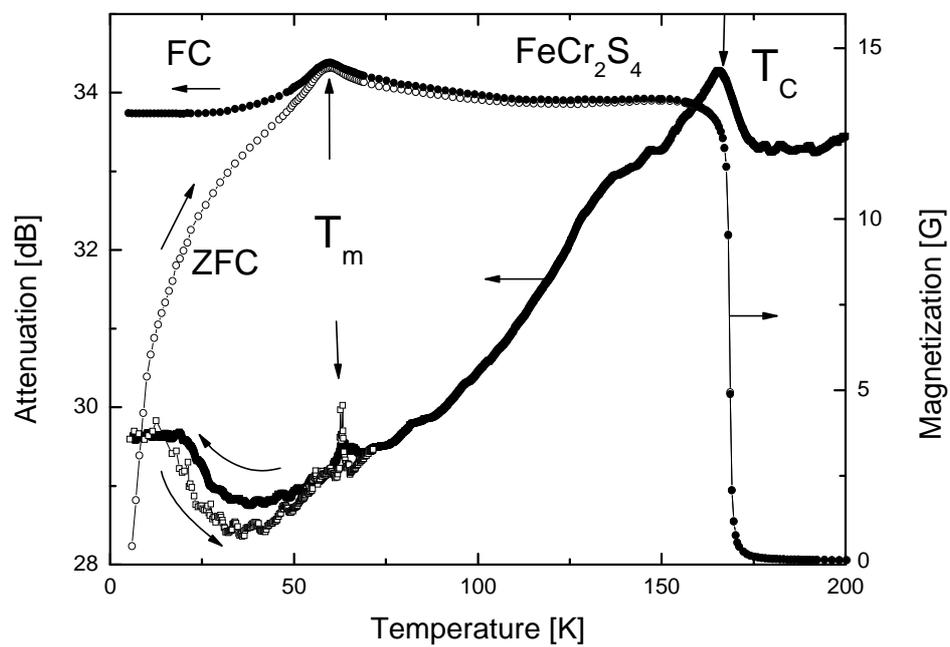

FIG.2 C. Müller, V. Zestrea, V. Tsurkan, et al.

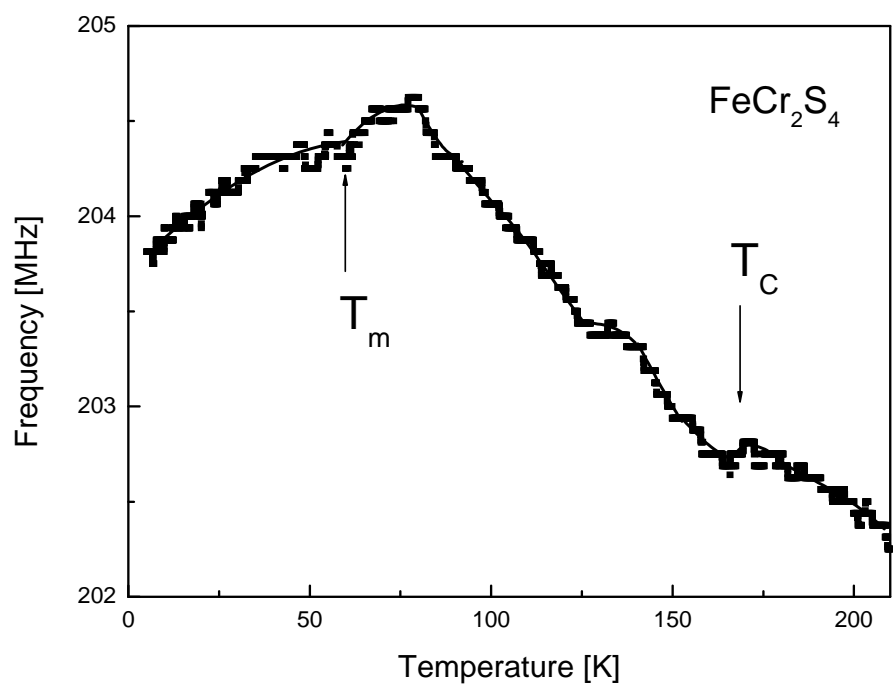

FIG.3  C. Müller , V. Zestrea, V. Tsurkan, et al.